\documentclass{phc-proc4-auth}

\usepackage{graphicx}
\usepackage{amssymb}

\begin{document}

\begin{frontmatter}


%\ead{email address}

\title{Intraplanar couplings in the CuO$_2$ lattice of cuprate 
superconductors}


\author{J. R\"ohler}
\address{Fachgruppe Physik, Universit\"at zu K\"oln, Z\"ulpicher Str. 77, D-50937 K\"oln, 
Germany }
\ead{abb12@uni-koeln.de}
\ead[url]{www.uni-koeln.de/$\sim$abb12}

\begin{abstract}
% Text of abstract
We have investigated the doping dependencies of the basal areas in
single-layer high-$T_{c}$ cuprates La$_{2-x}$Sr$_{x}$CuO$_{4}$ and HgBa$_{2}$CuO$_{x}$, as
well as in two-layer Y$_{1-y}$Ca$_{y}$Ba$_{2}$Cu$_{3}$O$_{x}$ and
 HgBa$_{2}$CaCu$_{2}$O$_{x}$.  The basal areas not only tend to shrink
on hole doping, as expected from single electron quantum chemistry,
but exhibit also a ``bulge'' around optimum doping.  We attribute the
``bulge'' to the effects of the strongly correlated quantum liquid on 
the CuO$_2$ lattice, rendering it nearly incompressible around 
optimum doping, but highly compressible in the weakly overdoped regime. 
Inhomogenous doping cannot account for this anomaly in the electronic
compressibility of the CuO$_2$ lattice.
\end{abstract}

\end{frontmatter}

% main text

%
\begin{figure}[b]

\includegraphics*[width=8.25cm]{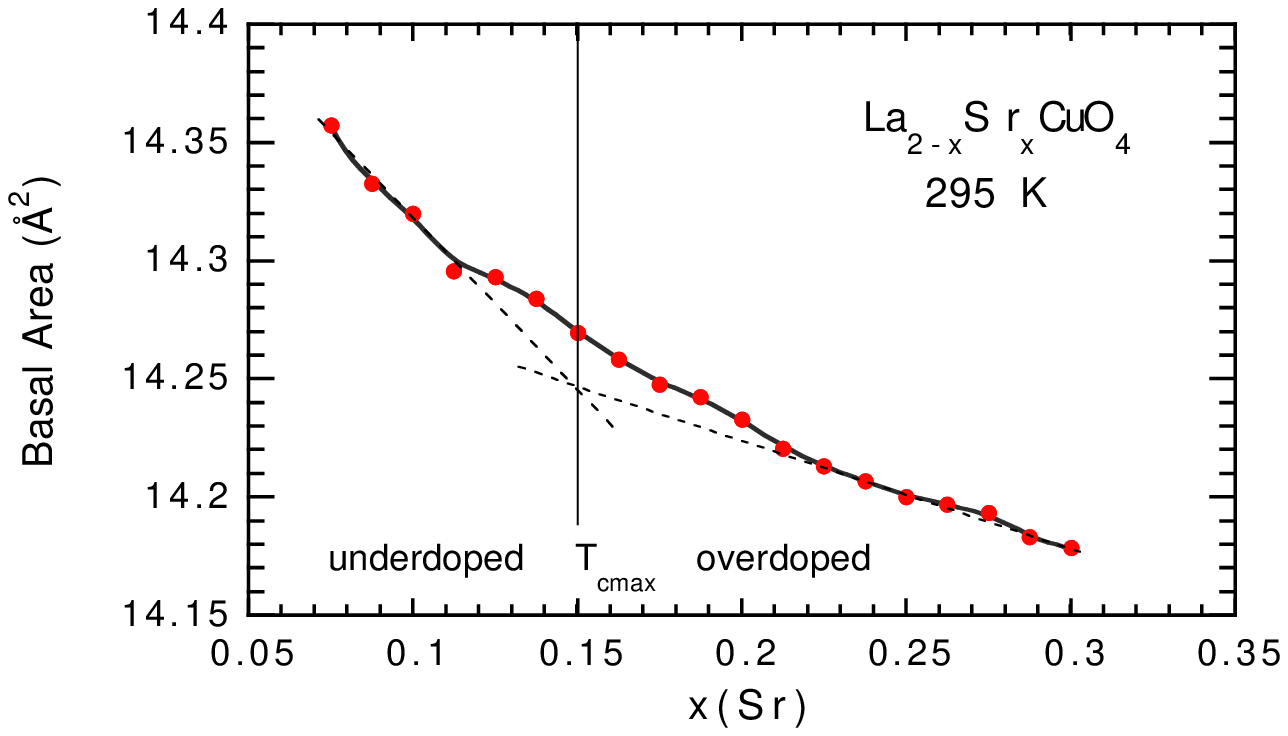}
\includegraphics*[width=7.5cm]{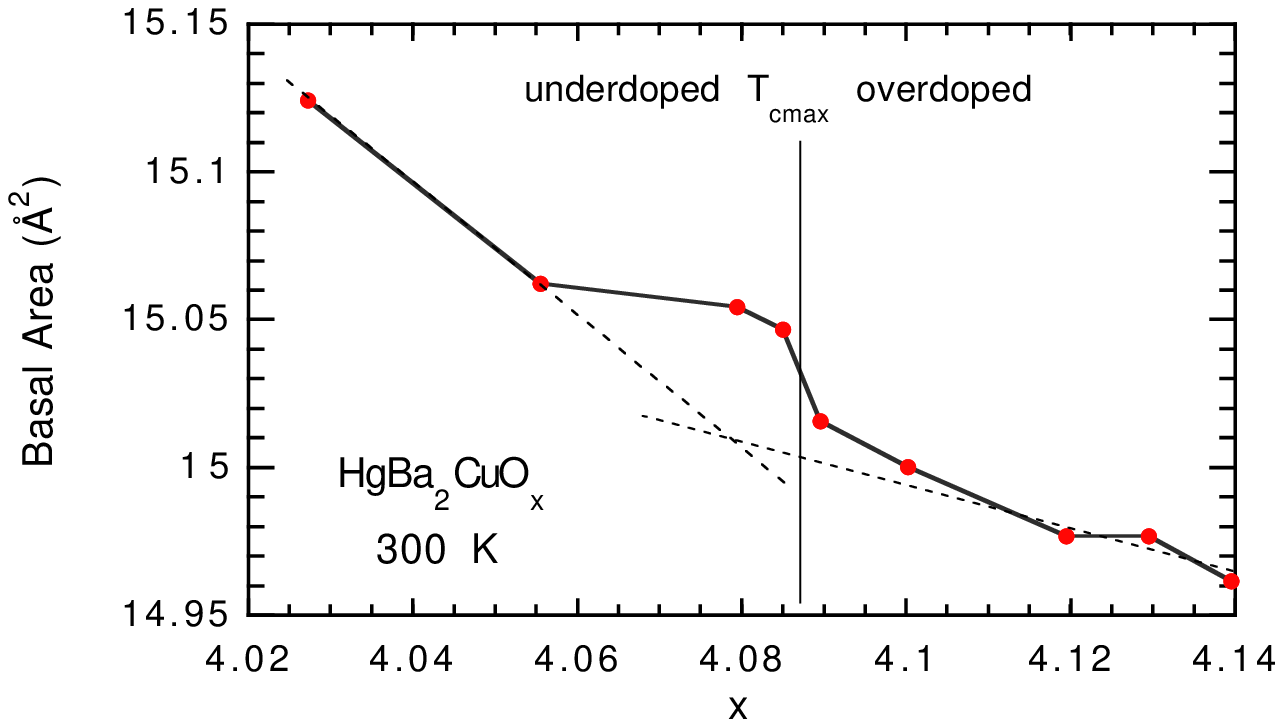}

\caption 
{Basal areas of one-layer cuprates as a function of doping.}
\end{figure}
The detailed geometry and the dimensions of the Cu--O layers in the
superconducting cuprates are controlled by extrinsic and intrinsic
effects.  The extrinsic effects arise mainly from the complex
stereochemistry of the structural blocks outside the planes doping the
holes into the Cu--O layers.  Mixed valent cations, oxygen defects,
disorder, structural misfit and strongly polarizing bonds induce
various types of strain, displacing the copper and oxygen atoms in-plane 
and out-of-plane.  The intrinsic effects are connected with the generic
properties of the quantum liquid created by the hole doping into a
Mott insulator \cite{Roe02}. In this short note we present experimental evidence
for that the strong electron-electron correlations inherent in this
quantum liquid affect significantly the area of the CuO$_2$ lattice.

Hole doping removes electrons from the antibonding
$\sigma^*$Cu$3d_{x^2-y^2}$O$2p_{x,y}$ band, increases the amount of
covalent character in the Cu--O bonds, and is thus expected to shorten
them. Strain from the chemically complex extrinsic properties, in
particular anisotropies, may however mask or even invert the expected
shortening of the individual bond lengths, $e.g.$ in
YBa$_{2}$Cu$_{3}$O$_{x}$ where the $a$\/-axis contracts as the
$b$\/-axis expands on hole doping \cite{Roe02,Kal01}.  
The basal area, $B$, defined by the square of the basal 
Cu--Cu distances, turns out to be almost unaffected by externally 
driven changes of the crystal symmetry, and hence 
allows for comparisons between various systems with very 
different doping chemistries.

Fig.  1 displays as a function of doping the basal areas of the
one-layer systems La$_{2-x}$Sr$_{x}$CuO$_{4}$ ($T_{cmax}=36$ K)
\cite{RadJor} and HgBa$_{2}$CuO$_{x}$ ($T_{cmax}=96$ K) \cite{FukTan},
Fig.  2 of the two-layer systems
Y$_{1-y}$Ca$_{y}$Ba$_{2}$Cu$_{3}$O$_{x}$ ($T_{cmax}=92$ K)
\cite{Kal01} and HgBa$_{2}$CaCu$_{2}$O$_{x}$ ($T_{cmax}=127$ K)
\cite{FukTan}.  $B$ is derived from the lattice parameters as reported
by Radaelli $et$ $al.$ \cite{RadJor} (LSCO), Fukuoka $et$ $al.$
\cite{FukTan} (HBCO, HBCCO), and Kaldis $et$ $al.$ \cite{Kal01}
(YBCO).  The thick drawn out lines connecting the data points are a
guide to the eye.  The overall behavior of $B(x)$ exhibits
surprisingly strong similarities in all systems under comparison: $i.$
As expected from the increaseing covalency with hole doping the basal
areas shrink appreciably (by about 1\%) between the insulator--metal
transition and the strongly overdoped regime.  $ii.$ $B(x)$ exhibits a
``bulge'' around optimum doping.  The ``bulge'' is weakest in
La$_{2-x}$Sr$_{x}$CuO$_{4}$ and strongest in
HgBa$_{2}$CaCu$_{2}$O$_{x}$.  The thin dashed straights are fitted to
the data points at the strongly under- and overdoped ends and,
extrapolated towards the underdoped-overdoped phase boundary, turn out
to intersect around optimum doping, $x_{opt}$.  We use them as rough
approximations defining the single electron quantum chemical
``background'' $B_{0}(x)$.  The change of slope around $x_{opt}$
indicates that the Cu--O bonding changes from a nearly ionic character
in the lightly doped regime to covalency in the strongly overdoped
regime.

Thus the ``bulge'' may be seen to ride on the single electron quantum
chemical background.  We attribute it to the effect of strong
electron-electron correlations on the CuO$_2$ lattice of the doped
Mott insulator.  Its basic physics may be described in terms of the 
competition between the magnetic exchange energy
$J$ and the kinetic energy per hole $xt$, underlying a
no-double-occupant constraint.  Here the doped holes appear as
vacancies in the background of a spin singlet liquid in the CuO$_2$
lattice. The strong repulsive interaction set by $J\simeq
1500$ K is expected to render the underlying lattice 
almost incompressible, at least much less 
compressible than a liquid of noninteracting holes residing in 
the same grid. Defining a electronic compressibility by
$\kappa_{e}\propto -\partial (B-B_{0})/\partial \mu$, where the 
pressure exerted by the holes, $\mu \propto xt$, we find
$\kappa_{e}\simeq 0$ close to optimum doping.  Apparently the strongly
correlated quantum liquid renders the lattice almost incompressible
where $T_{c}$ achieves its maximum.

The disappearence of the ``bulge'' 
in the weakly overdoped regime may be accompanied by a 
structural phase transformation of a martensitic type as recently 
reported for  YBa$_{2}$Cu$_{3}$O$_{x}$ \cite{{KalLoe},{Roe02}}.

We can however exclude that the ``bulge'' is connected with the formation of 
stripes or related nano-scale domains caused by inhomogenous doping. 
First, La$_{2-x}$Sr$_{x}$CuO$_{4}$, due to its relatively high 
structural compliance \cite{BilDux} $the$ system most susceptible 
to stable long range ordered stripes, exhibits the weakest ``bulge''. 
Second, the Hg-cuprates, due to their almost flat CuO$_2$ lattices 
(zero structural compliance) almost unable to accomodate stripes, 
exhibit the strongest ``bulges''. We conclude that homogenous doping governs the 
doping dependence of the basal area.

\begin{figure}[t]

\includegraphics*[width=8.25cm]{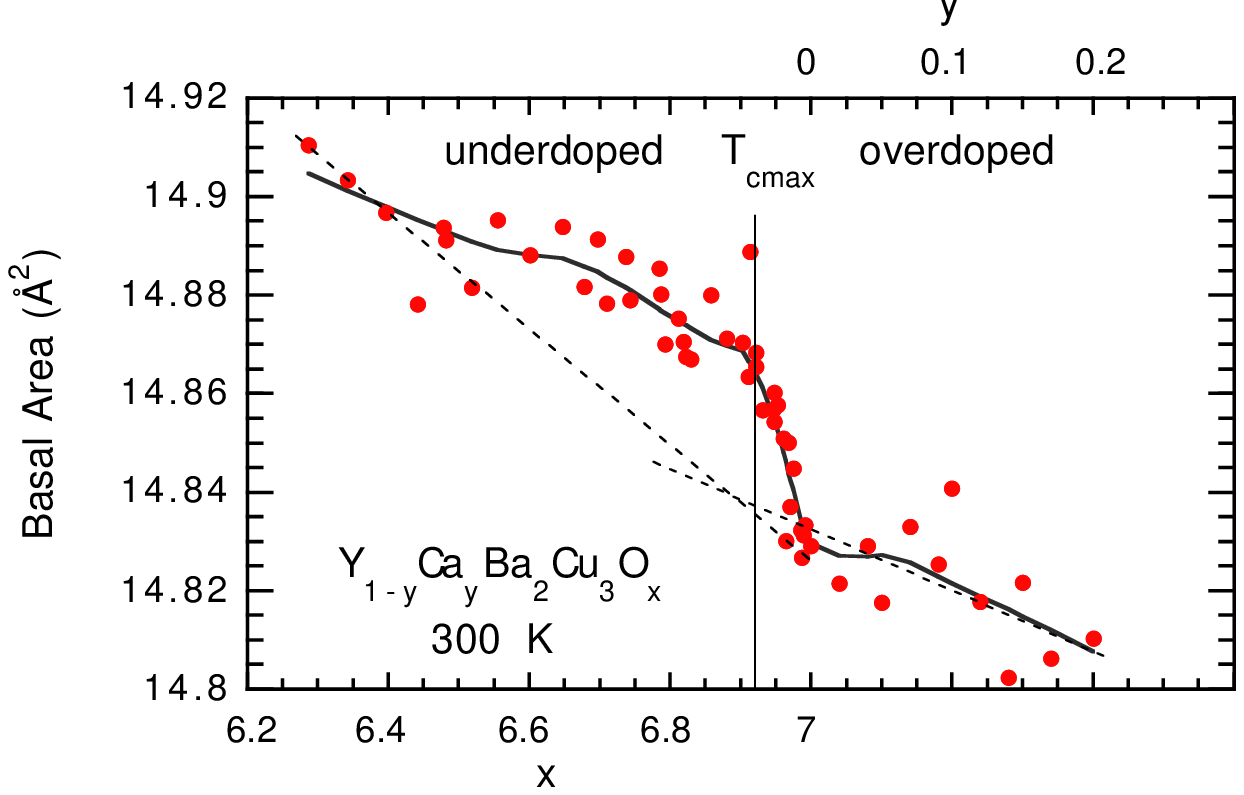}
\includegraphics*[width=8.25cm]{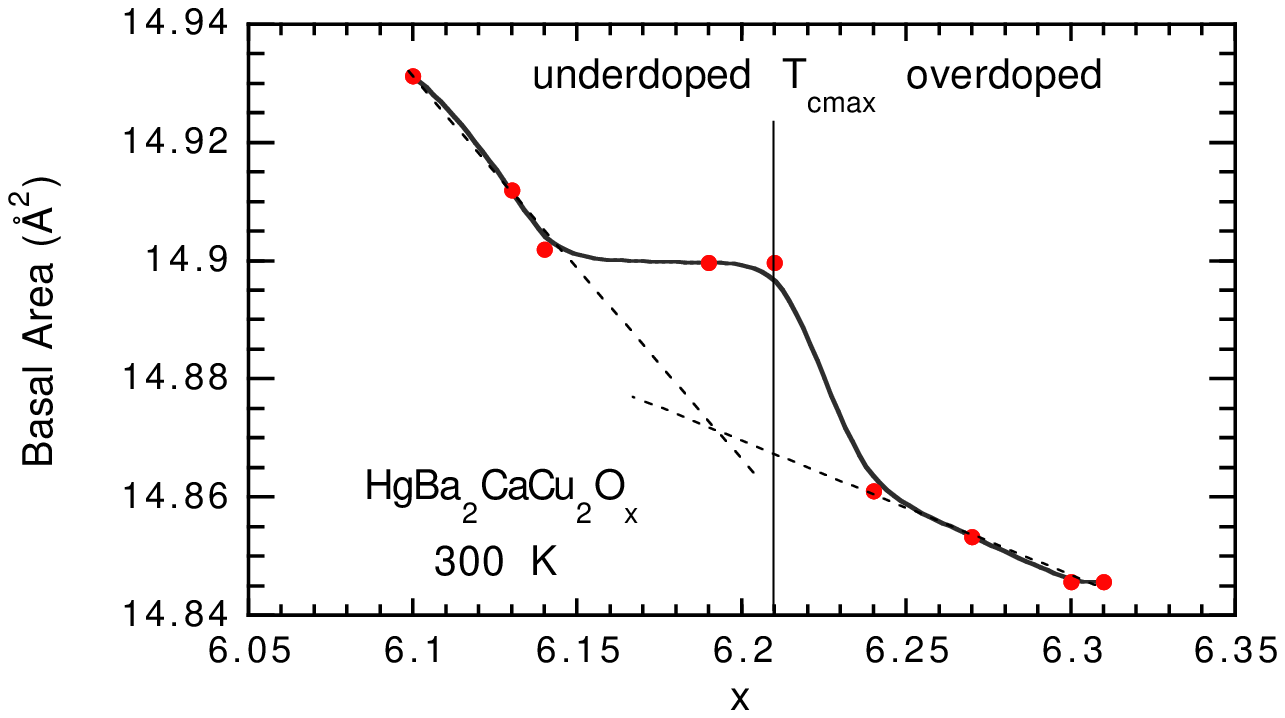}

\caption{Basal areas of two-layer cuprates as a function of 
doping.}\label{2-L}
\end{figure}

% \appendix

%\section{}
%\label{}

\end{document}